\documentclass[aps,prd,twocolumn,superscriptaddress]{revtex4-1} 
\usepackage{fullpage, amsmath, xparse, amssymb, mathtools, graphicx, braket,xcolor,verbatim}
\usepackage{natbib}
\usepackage[colorlinks=true,linkcolor=blue,citecolor=blue]{hyperref}

\newcommand{\mk}[1]{{\textcolor{green}{{\bf MK}: #1}}}
\newcommand{\aeber}[1]{{\textcolor{red}{{\bf AE}: #1}}}

\usepackage{multirow}
\usepackage{cancel}

\begin{document}
\title{When quantum corrections alter the predictions of classical field theory for scalar field dark matter}
\author{Andrew Eberhardt}
\email{Corresponding author. \\ aeberhar@stanford.edu}
\affiliation{Kavli Institute for Particle Astrophysics and Cosmology, Menlo Park, 94025, California, USA}
\affiliation{Physics Department, Stanford University, Stanford, California, USA}
\affiliation{SLAC National Accelerator Laboratory}
\author{Michael Kopp}
\affiliation{Stockholm University,
Hannes Alfv\'ens v\"ag 12, SE-106 91 Stockholm, Sweden}
\author{Tom Abel}
\affiliation{Kavli Institute for Particle Astrophysics and Cosmology, Menlo Park, 94025, California, USA}
\affiliation{Physics Department, Stanford University, Stanford, California, USA}
\affiliation{SLAC National Accelerator Laboratory}
\begin{abstract}

We investigate the timescale on which quantum corrections alter the predictions of classical field theory for scalar field dark matter. This is accomplished by including second order terms in the evolution proportional to the covariance of the field operators. When this covariance is no longer small compared to the mean field value, we say that the system has reached the ``quantum breaktime" and the predictions of classical field theory will begin to differ from those of the full quantum theory. While holding the classical field theory evolution fixed, we determine the change of the quantum breaktime as total occupation number is increased. This provides a novel numerical estimation of the breaktime based at high occupations $n_{tot}$ and mode number $N=256$. We study the collapse of a sinusoidal overdensity in a single spatial dimension. We find that the breaktime scales as $\log(n_{tot})$ prior to shell crossing and then then as a powerlaw following the collapse. If we assume that the collapsing phase is representative of halos undergoing nonlinear growth, this implies that the quantum breaktime of typical systems may be as large as $\sim 30$ of dynamical times even at occupations of $n_{tot}\sim 10^{100}$. 

\end{abstract}

\maketitle

\section{Introduction} \label{sec:Intro}
Discovering the properties of dark matter remains an outstanding problem in cosmology \cite{Peebles1933}. Observations indicate that the dark matter is cold and collisionless, though its specific particle nature is not known \cite{Spergel_2003, Aghanim2018}. Numerical studies allow us to compare observations with the predictions of simulations of different dark matter models \cite{Vogelsberger}. Understanding where our numerical methods are valid is then necessary to interpreting the predictions of these studies. 

One interesting dark matter model is scalar field dark matter (SFDM) \cite{Hu2000, Preskill:1982, Turner1983, Schive_2016, mocz2019, Guth2015, Seidel:1994zb, Arvanitaki_2020, ABBOTT1983133, Marsh_2016, hui2021wave, pozo2021detection, Lentz2019}. This model involves extremely light scalar particles with masses typically somewhere in the range $10^{-5} - 10^{-22} \, \mathrm{eV}$. The QCD axion is a potential SFDM candidate \cite{Preskill:1982} making simulations of this model relevant for many axion detection experimental setups \cite{admx, hui2021wave, Braine2020, Zhong2018, Salemi}. 

The low mass in this model creates a system where phase space occupation numbers are extremely high and, on the lowest mass end of the model, particles have a typical deBroglie wavelength on the $\sim \, \mathrm{kpc}$ length scale \cite{Hu2000, ABBOTT1983133, Guth2015, Arvanitaki_2020}. It is has been shown that power at length scales below the deBroglie wavelength would be suppressed but structure on larger scales would be left similar to standard $\Lambda$ cold dark matter cosmology ($\mathrm{\Lambda CDM}$) \cite{mocz2019, Hu2000, Arvanitaki_2020}. This model then allows one to tune the particle mass to account for discrepancies between simulation and observation on small scales, creating a potential solution to small scale structure problems. On small scales observations of structure deviate from the predictions of dark matter only cosmological simulations \cite{Weinberg2015, Arvanitaki_2020, Marsh_2016, Schive:2014dra, pozo2021detection}. Typically this deviation is summarized as three problems. The core-cusp problem: describing the observed flattening of galactic density profiles (cores) near the center of galaxies, as opposed to an increasingly steep density profile (cusp) \cite{navarro1996,Walker_2011}. The missing satellites problem: describing the lack of observed Milky-Way satellite galaxies \cite{Klypin:1999uc, Moore1994, Moore1999}. And the too-big-to-fail problem: describing the lack of observed massive dark matter sub halos \cite{Papastergis, Boylan-Kolchin2011}. This problem admits a number of solutions in addition to scalar field dark matter. Improved observational efforts, self interacting dark matter, and Baryonic effects are also suspected as possible solutions \cite{Governato2012, Nadler:2019zrb, vandenBosch:1999ka, Brooks_2013, Papastergis2016, Tulin2017}. It should also be mentioned that scalar field dark matter is associated with other interesting phenomenology in regions of parameter space where small scale structure is left largely unchanged \cite{Arvanitaki_2020, Marsh_2016, hui2021wave}.

The high occupation numbers of degenerate Bosons motivate approximating scalar field dark matter with a classical mean field theory (MFT) \cite{Guth2015, Kirkpatrick2020, Hertberg2016, mocz2019, Hu2000, Arvanitaki_2020, ABBOTT1983133, Marsh_2016, eberhardt2021Q}. Likewise, for dark matter models which have misalignment as their production mechanism, the initial quantum state is thought to be well described by a coherent state \cite{ABBOTT1983133, Preskill:1982}. Such models would therefore be well described at early times by MFT. Generally, numerical studies of SFDM proceed by solving the Schr\"odinger Poisson equations on a grid using a single classical field to represent the dark matter phase space \cite{mocz2019, Hu2000, Schive:2014dra, zhang2019}. It is known that MFT accurately approximates weakly interacting highly degenerate Bosonic systems such as Bose Einstein Condensates and coherent electromagnetic radiation \cite{bial1977, Gross, Pi, BECrev}.

However the extent to which MFT can approximate the evolution of strongly interacting systems is a topic of interest \cite{Dvali_2018, Hertberg2016, Sikivie:2016enz, Dvali:2017eba, Dvali:2013vxa, sreedharan2020, Chakrabarty:2017fkd, kopp2021nonclassicality, Lentz2019, Lentz2020, eberhardt2021Q}. It is possible that the mean field description of SFDM fails in some regimes or on some timescale relevant to cosmological simulations and that quantum effects begin to become important \cite{Sikivie:2016enz, Seidel:1994zb, chakrabarty2021, Chakrabarty:2017fkd, kopp2021nonclassicality, Lentz2019}. This is due to the spreading of the quantum wavefunction around the classical solution, an effect generic to nonlinear systems \cite{Yurke1986, sreedharan2020, Lewenstein1996, Cabellero2008, eberhardt2021Q, eberhardt2021}. The timescale on which the classical and quantum evolutions begin to differ is often referred to as the ``quantum breaktime". Estimations of this timescale have been performed for scalar field dark matter \cite{Dvali_2018, kopp2021nonclassicality, Sikivie2012}. However, these estimations have either relied on significant approximations about the behavior of the spread of the wavefunction or simulations of significantly less complicated systems. The main focus of this work will be to characterize the behavior of quantum corrections for large systems undergoing gravitational collapse. We use a solver which tracks the leading order correction to the MFT which is proportional to the second order field moments. This allows us to directly numerically integrate the evolution of the quantum corrections in a way numerically efficient enough to allow us to estimate the behavior of these corrections for systems much larger than have been previously investigated numerically. 

This is done using the field moment expansion method (FME), described in \cite{eberhardt2021}, which estimates the relative size of quantum correction terms in the evolution of a system initially well described by mean field theory. When these terms pass a certain threshold we define MFT to have broken down. We estimate the quantum breaktime as a function of occupation number and self interaction strength for the gravitational collapse of an initial spatial overdensity. An appendix contains a brief discussion of other systems we tested.

The paper is organized as follows. In Section \ref{sec:background}, we discuss background on mean field theory, its quantum deviations and corrections. Section \ref{sec:numerics} explains the numerical implementation of the various theories. Section \ref{sec:results} contains results of applying this method to scalar field dark matter. We include a general discussion of the results in Section \ref{sec:conclusions}, focusing on the implication for simulations of scalar field dark matter.
\section{Background} \label{sec:background}

\subsection{Quantum description}

The evolution of a quantum system is determined by its Hamiltonian and the initial conditions of the quantum state. In this work we will use the following Hamiltonian

\begin{equation} \label{Ham}
    \hat H = \sum_j^M \omega_j \hat a_j^\dagger \hat a_j + \sum_{ijkl}^M \frac{\Lambda_{kl}^{ij}}{2} \hat a_k^\dagger \hat a_l^\dagger \hat a_i \hat a_j \, .
\end{equation}

Which, with appropriate choice of $\omega_j$ and $\Lambda^{ij}_{kl}$, describes a variety of non-relativistic field physics, see for example \cite{Sikivie:2016enz, Sikivie2012, Hertberg2016, eberhardt2021Q}. $M$ is the total number of allowed modes, $\hat a_j$ and $\hat a_j^\dagger$ are annihilation and creation operators on mode $j$ respectively, and $\omega_j$ is the kinetic energy associated with mode $j$. Throughout this paper we will assume that the mode functions are the momentum eigenstates and that $\omega_j \propto \hbar j^2 / (2m)$. The matrix $\Lambda^{ij}_{kl}$ describes the potential energy of the field, including self interactions. If we write this matrix as 

\begin{equation} \label{Lambda}
    \Lambda^{ij}_{kl} = \left( \frac{C}{2(p_k - p_i)^2} + \frac{C}{2(p_k - p_j)^2}\right) \delta^{ij}_{kl}  
\end{equation}

then $C = -4 \pi\, G\, m^2/ \hbar $ describes the gravitational self interaction and $C = 4 \pi k_e\,q\, m/ \hbar $ describes the electrostatic self interaction, via four point vertex. 

In the Heisenberg picture, the dynamical variables are the operators themselves, which evolve according to the Heisenberg equation of motion

\begin{align} \label{eqnMotion}
    \partial_t \hat a_p &= i [\hat H, \hat a_p] = -i\left[ \omega_p \hat a_p + \sum_{ijl} \Lambda^{ij}_{pl} \hat a_l^\dagger \hat a_i \hat a_j \right] \, .
\end{align}

All that is left is to specify the initial quantum state. Because we are primarily concerned with ultra light scalar field dark matter we will assume that the initial conditions are approximately coherent. We will define what we mean by approximately coherent in later sections. For now we will say that is that the initial quantum state is close to 

\begin{equation} \label{coherentStates}
    \ket{\Vec{z}} = \bigotimes_{i=1}^M \exp \left[ -\frac{|z_i|^2}{2} \right] \sum_{n_i=0}^\infty \frac{ z_i^{n_i}}{\sqrt{n_i!}} \ket{n_i} \, .
\end{equation}

where $\Vec{z} \in \mathbb{C}^M$ parameterizes the state. Physically a coherent state is defined ss a displacement from the vacuum state. $\ket{n_i}=\frac{ (\hat a^\dagger_i(t=0))^{n_i}}{\sqrt{n_i!}} \ket{0}$ represents the number eigenstates corresponding to having $n_i$ particles in the $i$th mode.  For axions produced by the misalignment mechanism a coherent state is thought to initially provide an accurate description \cite{ABBOTT1983133, Preskill:1982}.

Coherent states have the following important property: for a coherent state the expectation value of any normally ordered operator is given

\begin{align} \label{coherentProp}
    \braket{\Vec{z} | \, f(\set{\hat a^\dagger}) \, g(\set{\hat a}) \, | \Vec{z} } = f(\set{z^\dagger}) \, g(\set{z})
\end{align}
where $\hat a = \hat a(t=0)$. 

\subsection{Obtaining the classical theory}

MFT is obtained by replacing the operators in equation \ref{eqnMotion} with complex numbers which are then the ``classical field". This means that the classical field equation of motion is written  

\begin{align} \label{clEqnMotion}
    \partial_t a^{cl}_p = -i\left[ \omega_p a^{cl}_p + \sum_{ijl}\Lambda^{ij}_{pl} a^{cl\dagger}_l a^{cl}_i a^{cl}_j \right] \, .
\end{align}

Typically, initially, $a^{cl}_p$ corresponds to the mean of the field operator $\hat a_p$, that is 

\begin{align} \label{clEqnMotionInitialCons}
    a^{cl}_p \biggr\rvert_{t=0} = \braket{\hat a_p} \biggr\rvert_{t=0} \, .
\end{align}

And in fact we can think of the MFT equation of motion as resulting from the following approximation

\begin{align} \label{clApprox}
    \partial_t \braket{\hat a_p} = \braket{F(\set{\hat a})} \approx F(\set{ \braket{\hat a} }) \, .
\end{align}

Where $F(\set{\hat a})$ is the right hand side of equation \ref{eqnMotion}. The first equality above is the result of taking the expectation value of equation \ref{eqnMotion}. The approximation gives equation \ref{clEqnMotion}.

This approximation is accurate when the mean value of the operator is large compared to its root variance. We will make this requirement more precise in the next section. 

For coherent state initial conditions, at $t=0$ we can use the property in equation \ref{coherentProp} to restore the approximation in equation \ref{clApprox} to an equality. 

Notice also that taking a Fourier transform of equation \ref{clEqnMotion} and defining $\Lambda^{ij}_{pl}$ as we have in the previous section recoveres the familiar Schr\"odinger-Poisson equations of motion. 

\begin{align} \label{Schr_eqn}
    \partial_t \psi(x) &= -i \left[ \frac{-\nabla^2}{2m}  + m  V(x)  \right] \psi(x) \, , \\ 
    \nabla^2  V &= 4 \pi G m \, \psi^\dagger(x)  \psi(x) \, , .
\end{align} 

where

\begin{equation} \label{psi2a}
    \psi(x) =  \sum_i a_i^{cl} u^\dagger_i(x) \,
\end{equation}

$u_i(x)$ is the momentum eigenstate with momentum $p_i$. Note the above equations describe gravity, in the electromagnetic case we simply make the replacements $G \rightarrow -k_e$ and $\sqrt{m} \rightarrow \sqrt{q}$. 

\begin{widetext}

\subsection{Corrections to the classical theory} \label{sec:correctionsToClassical}

In order to understand how the corrections to the classical theory arise we return to equation \ref{clApprox}. This approximation can be returned to an equality if we write the rightmost side as the following series 

\begin{align} \label{eqMotion}
    \partial_t \braket{\hat a_p} &= \braket{F(\set{\hat a})} \nonumber \\
    &= F(\set{ \braket{\hat a} }) + \left(  \sum_{\set{\hat a}} \braket{\delta \hat a_i \delta \hat a_j} \frac{\partial^2}{\partial \hat a_i \partial \hat a_j }  + \braket{\delta \hat a_i \delta \hat a^\dagger_j}  \frac{\partial^2}{\partial \hat a_i \partial \hat a^\dagger_j }  \, + \dots \right) F(\set{ \braket{\hat a} }) .
\end{align}

The series in the parentheses represent how higher order moments effect the exact evolution of the mean field. Notice that each term in the sum is written as a central moment weighting a derivative of the same order representing the response of the evolution function to that central moment. This kind of expansion is commonly used to approximate probability distributions. 
\end{widetext}

By applying this expansion to the evolution of the expectation value of any operator we can find quantum correction terms to the classical equations of motion. Notice that at each order in this process we introduce additional moments that themselves have equations of motion. This process then generates an infinite series of coupled equations of motion. If the initial conditions and evolution are such that the growth of each central moment of a given order is hierarchical, i.e. moments of a higher order grow slower than ones of a lower order, than we can truncate this series at some order and always improve upon the accuracy of a lower order truncation up until the highest order correction terms become large. For coherent state initial conditions we expect such a hierarchical growth condition to be satisfied. 

Very approximately the ratio of the second order and classical terms can be given by the following parameter 

\begin{equation} \label{eqQ}
    Q \equiv \sum_i^M \frac{\braket{\delta \hat a_i^\dagger \delta \hat a_i}}{n_{tot}} \, .
\end{equation}
For a coherent state this parameter is $0$. When $Q$ is no longer much smaller than $1$ we expect the quantum correction terms to begin becoming important and for the classical theory to break down. Therefore, the quantum breaktime will be on the same timescale. Using our hierarchical growth assumption we can also say that the evolution of the second order equations of motion will remain accurate until this time. Therefore, a second order expansion should be sufficient to predict the breaktime. 

We therefore define a quantum breaktime as follows 

\begin{equation} \label{breakTime}
    Q(t_{br}) \equiv 0.15 \, .
\end{equation}

Note that our results should be fairly robust to the specific choice of $Q(t_{br})$, we choose $0.15$ here simple because it is clear that at this time it is likely true that the classical solution admits terms that are no longer sub-leading order and that the second order solution is still an accurate approximation of the exact quantum evolution, as the third order terms should still be sub-leading order. If the quantum correction terms are no longer small compared to the classical terms it is unlikely that the predictions of the classical field theory will remain accurate.

\begin{widetext}

\section{Numerical Implementation} \label{sec:numerics}
 
We solve the coupled equations of motion for the first and second order moments of the field operator. 
 
 \begin{align} \label{FME_eqn}
    \partial_t \braket{\hat a_p} &\approx -i\left[ \omega_p \braket{\hat a_p} + \sum_{ijl} \Lambda^{ij}_{pl} \left( \braket{\hat a_l^\dagger} \braket{\hat a_i} \braket{\hat a_j}  + \braket{\delta \hat a_i \delta \hat a_{j}} \braket{\hat a^\dagger_l} + \braket{\delta \hat a^\dagger_l \delta \hat a_{i}} \braket{\hat a_j} + \braket{\delta \hat a^\dagger_l \delta \hat a_{j}} \braket{\hat a_i} \, \right) \right] \, , \\
    \partial_t \braket{\delta \hat a_i \delta \hat a_j} &\approx -i \left[(\omega_i+\omega_j)\braket{ \delta \hat a_i  \delta \hat a_j} + \sum_{kp} \Lambda^{ij}_{pk} \braket{\hat a_k} \braket{\hat a_p}  \right. \label{daa_dt} \\
    &\left. + \sum_{kpl} \Lambda^{ij}_{pl} \left(  \braket{ \delta \hat a_l \delta \hat a_j} \braket{\hat a^\dagger_k} \braket{\hat a_p} + \braket{ \delta \hat a_j  \delta \hat a_p} \braket{\hat a^\dagger_k} \braket{\hat a_l} + \braket{\delta \hat a_k^\dagger \delta \hat a_j} \braket{\hat a_l} \braket{\hat a_p} \right) + (i \leftrightarrow j) \right]  \, , \nonumber \\
    \partial_t \braket{\delta \hat a_i^\dagger \delta \hat a_j} &\approx i \left[ (\omega_i - \omega_j)\braket{\delta \hat a_i^\dagger a_j} +  \right. \label{dba_dt}  \\
    &\left.  + \sum_{kpl}  \Lambda^{ij}_{pl}  \left( \braket{\delta \hat a_j \delta \hat a_k} \braket{\hat a^\dagger_p} \braket{\hat a^\dagger_l} + \braket{\delta \hat a^\dagger_p \delta \hat a_j} \braket{\hat a^\dagger_l} \braket{\hat a_k} + \braket{\delta \hat a^\dagger_l \delta \hat a_l} \braket{\hat a^\dagger_p} \braket{\hat a_k} \right) + (c.c., \, i \leftrightarrow j) \right] \, . \nonumber
\end{align}
 
 To do this we use the the CHiMES repository available at \href{https://github.com/andillio/CHiMES}{https://github.com/andillio/CHiMES}. These operators can then be used to estimate the breaktime as we have defined it in equation \ref{breakTime}. Appendix \ref{appendixB} contains a discussion of the choice of the mode number $N$ and timestep $dt$. 

\end{widetext}

\begin{figure*}[!ht]
	\includegraphics[width = .97\textwidth]{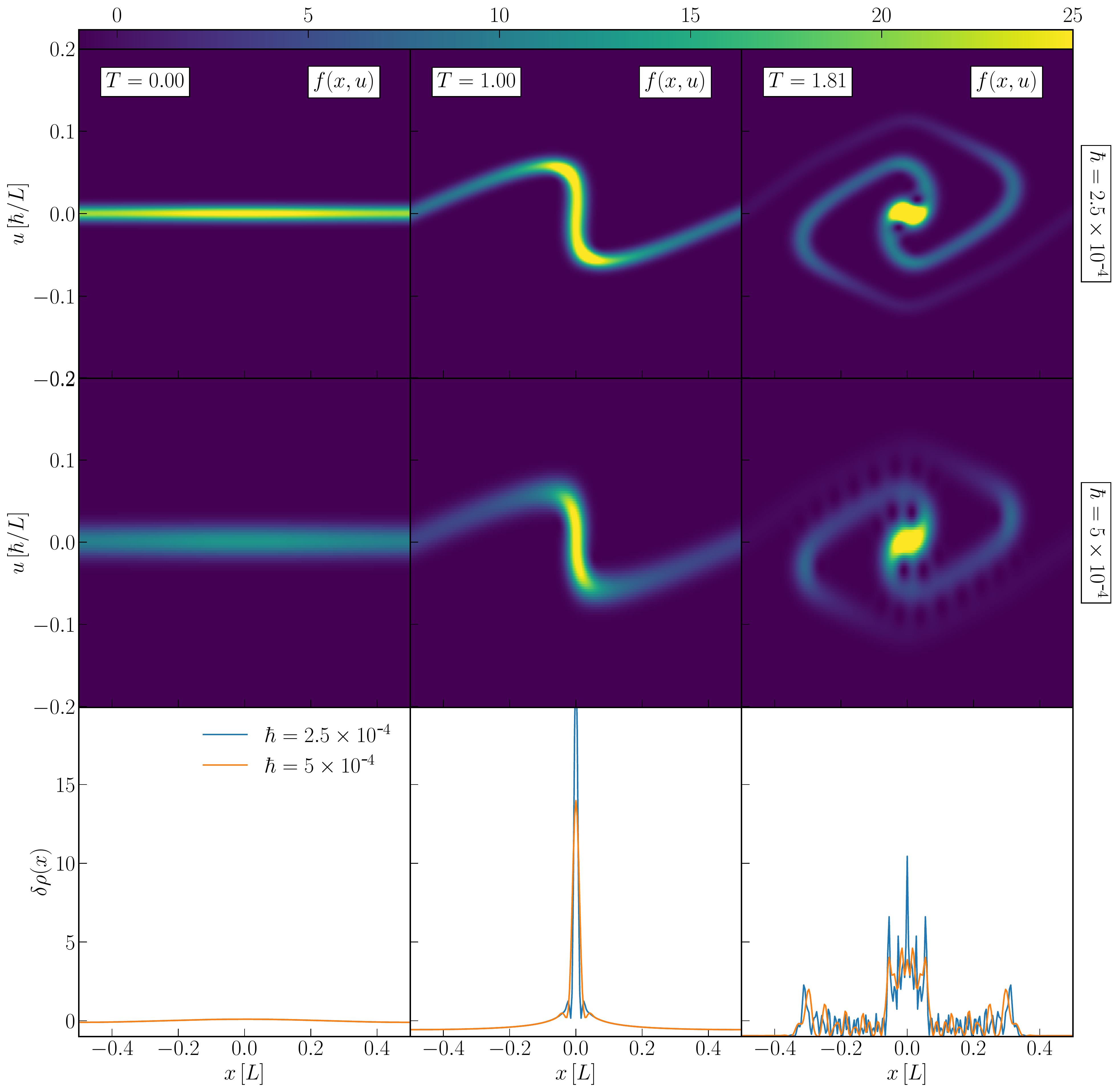}
	\caption{ Here we plot the evolution of sinusoidal over-density. The top two rows show the phase space density evolution for two different values of $\tilde \hbar$. The phase space is represented by the Husimi function of the classical field with fixed $\sigma_x$. The bottom row shows the density for both. Each column represents a different timestep, with the left most column showing the initial conditions, and the middle column showing shell crossing, the point at which the scaling of the breaktime transitions from logarithmic to a power law. }
	\label{fig:phaseSpaceSin}
\end{figure*}

\begin{figure}[!ht]
	\includegraphics[width = .49\textwidth]{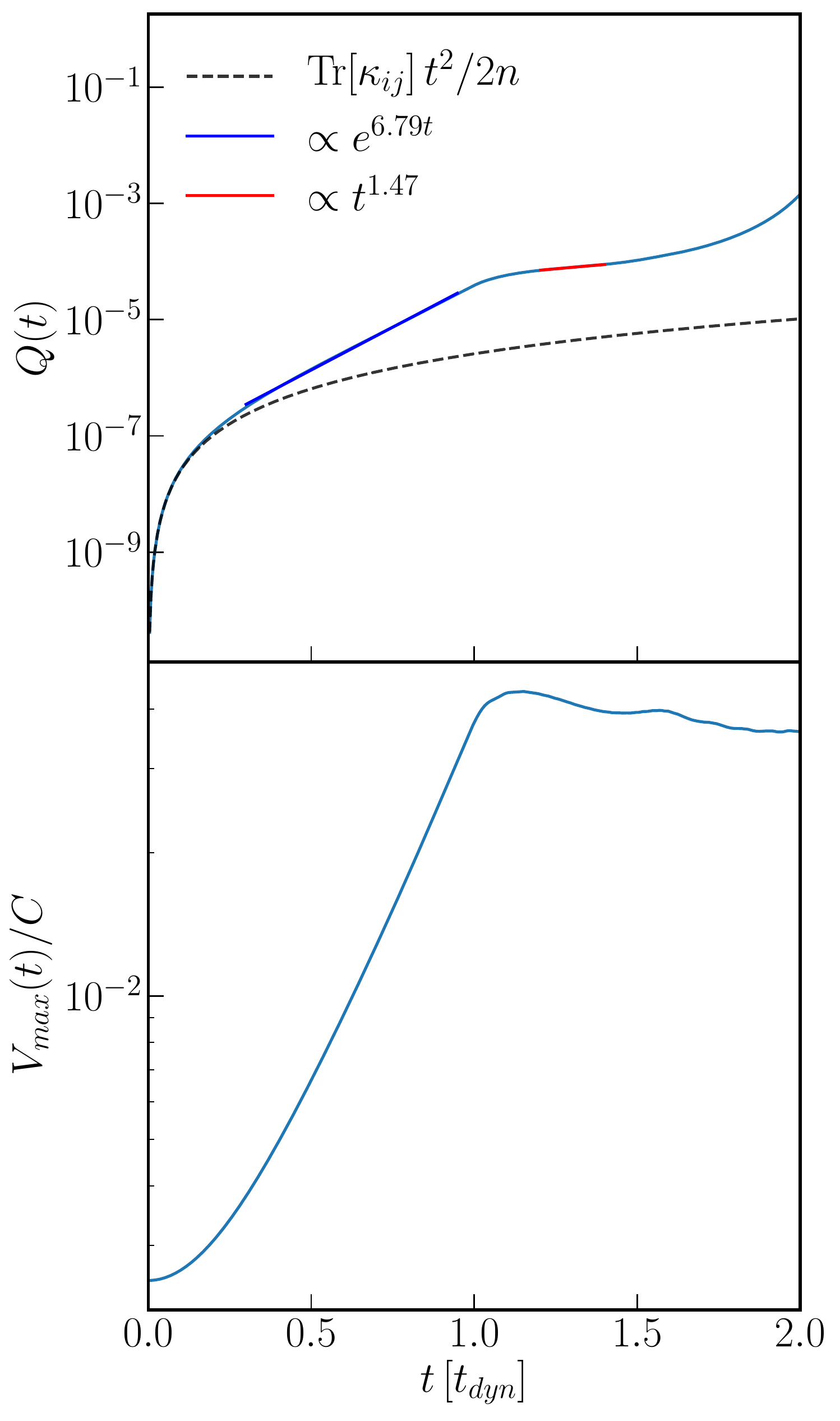}
	\caption{Here we plot how the wavefunction spreads at different stages in the evolution of the system, for the gravitational collapse of a sinewave overdensity. The top plot shows the evolution of $Q(t)$ and the bottom plot the maximum value of the potential, $V(t)$. We see that $Q$ first grows quadratically at early times when the growth of the potential is approximately linear, then begins to grow exponentially until the collapse time at $t= t_{dyn}$, after this time, it again grows according to a power law before undergoing runaway growth near the breaktime shortly after $t \approx 2 \, t_{dyn}$. The exponential growth of $Q$ corresponds to the exponential growth of the overdensity, following the collapse the wavefunction spreads much slower. Here we set $\tilde \hbar = 2.5 \times 10^{-4}$, $n_{tot} \approx 10^{10}$. }
	\label{fig:Qgrowth}
\end{figure}

\begin{figure*}[!ht]
	\includegraphics[width = .95\textwidth]{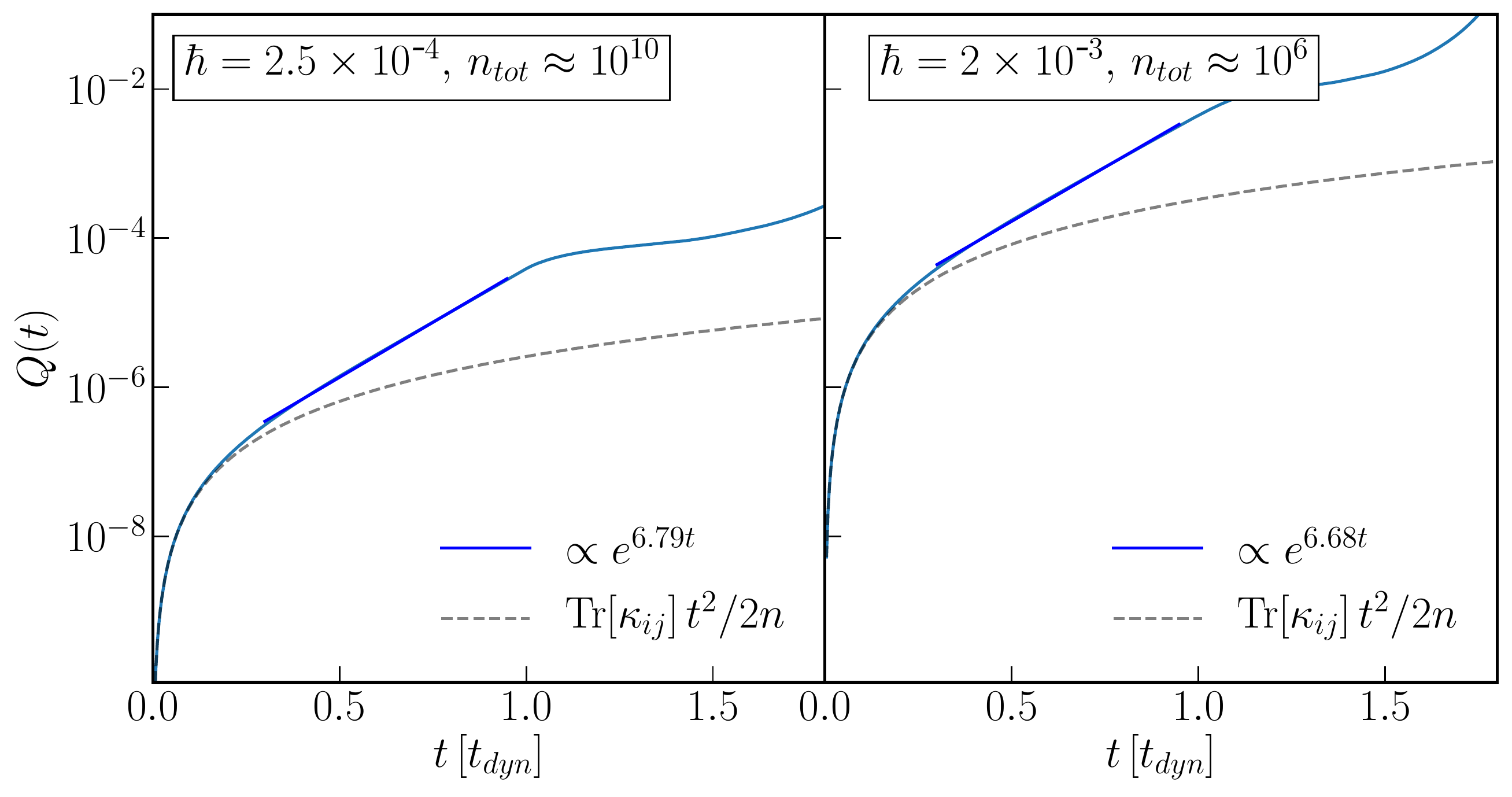}
	\caption{ Here we plot the evolution of $Q(t)$ for two systems with different $\tilde \hbar$ and $n_{tot}$. Notice that the behavior of $Q(t)$ is mostly constant between the two simulations despite a relatively large change in the simulation parameters. Both are fit well by the quadratic approximation at early times and then have a similar exponential growth prior to collapse at $t=t_d$. This is somewhat unsurprising as prior to shell crossing both systems have a similar mean field evolution, which is well approximated by classical particles and so we would not expect the evolution to depend sensitively on its scalar field qualities. }
	\label{fig:Qcompare}
\end{figure*}

\begin{figure*}[!ht]
	\includegraphics[width = .97\textwidth]{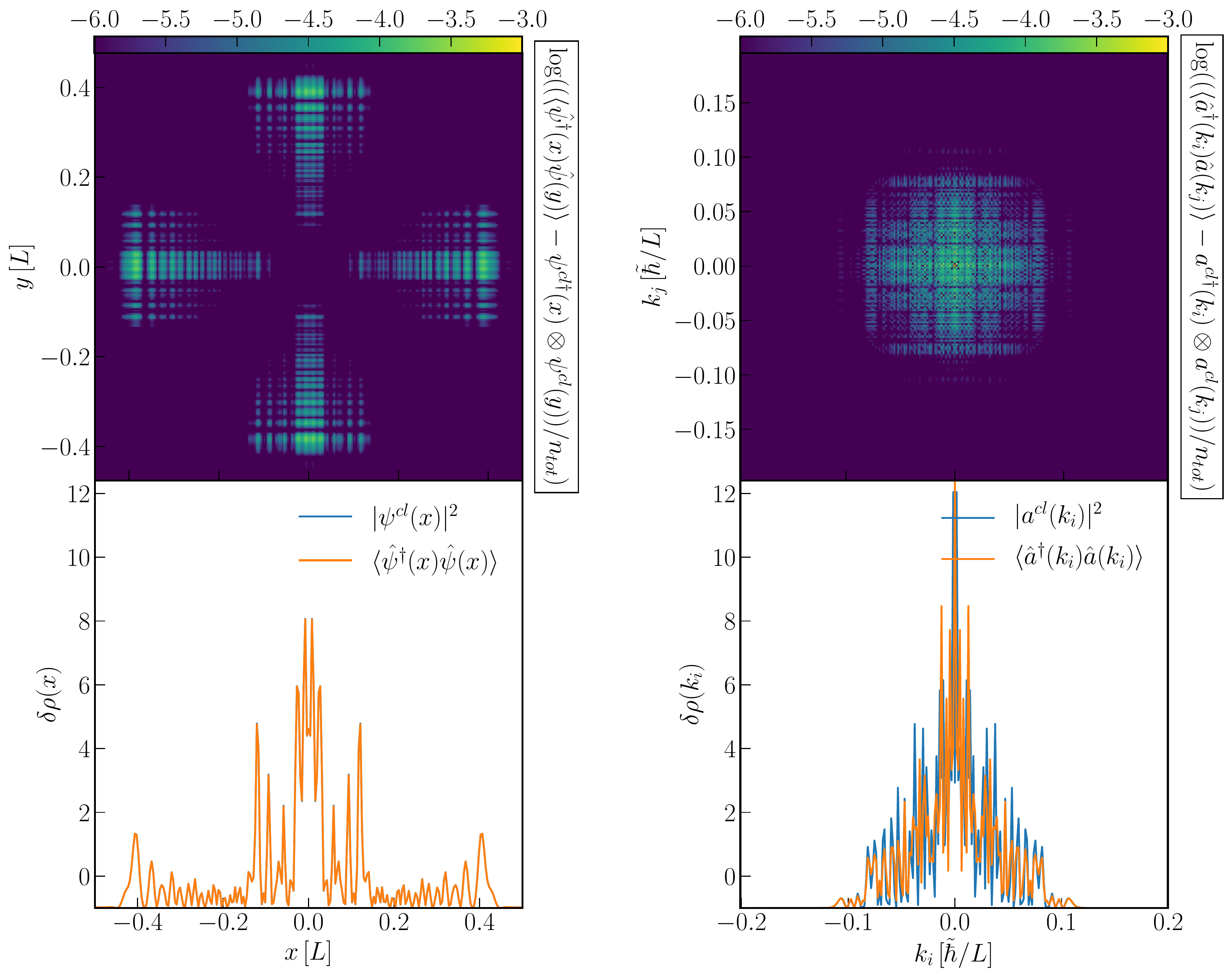}
	\caption{Here we look at the mean field theory differs from the field moment expansion, the left column showing spatial operators and the right momentum space operators. The top row shows $(\braket{\hat \psi^\dagger(y) \hat \psi(x)} - \psi^{cl \dagger}(y) \otimes \psi^{cl}(x))/n_{tot}$ and $(\braket{\hat a^\dagger(k_j) \hat a(k_i)} - a^{cl \dagger}(k_j) \otimes a^{cl}(k_i))/n_{tot}$. The bottom row shows the difference between the overdensities predicted by the number operators $\braket{\hat \psi^\dagger(x) \hat \psi(x)}$ and $\braket{\hat a^\dagger(k_i) \hat a(k_i)}$, and the square of the classical field $|\psi^{cl}(x)|^2$ and $| a^{cl}(k_i)|^2$. The difference between the blue and orange lines in the bottom row plots is given by the diagonal of the top row plots. We can see that quantum corrections occur in the momentum density but are off diagonal for the spatial covariance. This follows from the fact that the non-linearity is in the spatial potential and therefore directly effects the momentum evolution. }
	\label{fig:varCompare}
\end{figure*}

\begin{figure}[!ht]
	\includegraphics[width = .47\textwidth]{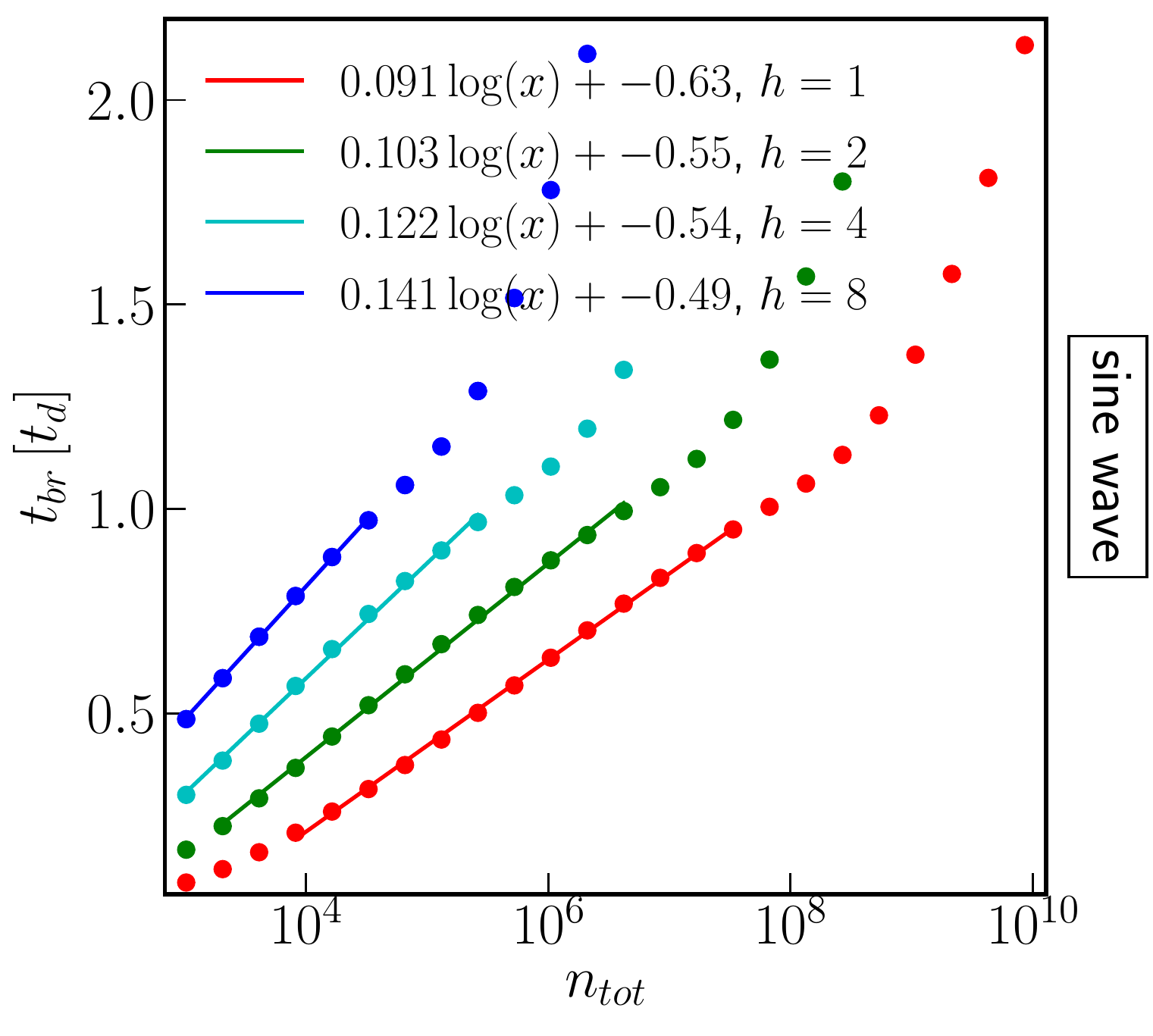}
	\caption{ Here we show the scaling of the breaktime, $t_{br}$, with total particle number, $n_{tot}$, for the sinusoidal overdensity. The breaktime is defined using $Q$, see equations \eqref{eqQ} and \eqref{breakTime}, which compares the spread of the wavefunction to the total number of particles. We can see that for systems with breaktimes shorter than the collapse time, i.e. $t_{br} < t_{d} $, the scaling of the breaktime with total occupation number is $\propto \log(n_{tot})$. However, after the collapse time, $t_{d}$, the scaling follows a power law. Each color line corresponds to a different classical field evolution, which is determined by the value of $\tilde \hbar$, we set $\tilde \hbar = h \times 2.5 \times 10^{-4}$. The trend of larger $\tilde \hbar$ corresponding to longer breaktimes implies that the breaktime decreases as mass is increased.}
	\label{fig:sinCollapse_tbr}
\end{figure}

\section{Results} \label{sec:results}

\subsection{Test problem set up}

Here we show the breaktime calculated for the collapse of a gravitational overdensity. We simulate the first and second order moments of the field using the method described in section \ref{sec:numerics}. The field moments are initialized as follows 

\begin{align}
    \psi(x) &= \sqrt{1 + \delta \, \cos \left( 2 \pi x / L \right)} / \mathrm{Norm} \, , \\
    \braket{\delta \hat a_i \delta \hat a_j} &= 0 \, , \\
    \braket{\delta \hat a^\dagger_i \delta \hat a_j} &= 0 \, .
\end{align}

Recall that $\psi(x)$ and $z_p$ are related via Fourier transform, i.e. $\psi(x) = \sum_p z_p u_p(x)$. We normalized each initial conditions such that the square norm equals the total number of particles, i.e. $\sum_p |z_p|^2 = \sum_x |\psi(x)|^2 = n_{tot}$. The overdensity tends to grow exponentially until shell crossing, see bottom plot in figure \ref{fig:Qgrowth}, resulting in a characteristic phase space spiral, see figure \ref{fig:phaseSpaceSin}. 

When simulating classical particles, for cold initial conditions, the growth of structure is ``scale free". This means that characteristic phase space spiral will continue indefinitely resulting in ever smaller structure. If we instead use a classical field, phase space structures smaller than $\hbar$ will be washed out, and a central core will form in the density as opposed to a cusp, see for example \cite{Eberhardt2020, Hu2000, hui2021wave}. The velocity dispersion of the collapsing system is set by the gravitational potential, which is well measured in cosmology. This velocity scale then sets a deBroglie wavelength which washes out structures below that spatial scale, this can be seen in figure \ref{fig:phaseSpaceSin}.

We will be interested in how the leading order and correction terms compare throughout the evolution as we scale the total number of particles, $n_{tot}$. The MFT corresponds to the infinite particle limit, thus as we scale the number of particles we will do so in a way that keeps the classical field evolution of the system fixed so that systems at different occupation can be meaningfully compared to one another.  This means that we hold $M_{tot}, \,  \tilde \hbar \equiv \hbar / m ,$ and $\Lambda \,  n_{tot}$ constant. Notice that for the numerical implementation purposes this implies the following scalings $m \propto 1/n_{tot}$, $\hbar \rightarrow 1/n_{tot}$, and $\Lambda \propto 1/n_{tot}$. We simulate each system until the breaktime, defined to occur at $Q(t_{br}) = 0.15$. Recall that $\hbar$ sets the scale for quantum corrections. When we perform the varying procedure described here we can parameterize the size of quantum corrections holding the mean field theory fixed such that their behavior can be studied as a function of the total particle number, and extrapolated to higher occupations.

\subsection{Spread of the wavefunction}

The wavefunction spreads around the mean field value throughout the evolution. This spread can be characterized by the growth of $Q$, defined in equation \eqref{eqQ}. This parameter measures how the mode variances compare with the total occupation number. We plot the growth of $Q$ for a highly occupied system with $n_{tot} \sim 10^{10}$ particles and $\tilde \hbar = 2.5 \times 10^{-4}$ in figure \ref{fig:Qgrowth}. The system is initialized so that at the initial conditions $Q(t=0) = 0$, representing an initial coherent state. Generally, we can characterize the growth of $Q$ by discussing four phases: initial quadratic growth, the collapsing phase, the post collapse phase, and runaway growth. It is useful to also look at the behavior of the overdensity, which can help describe the transition between these phases. We plot both $Q(t)$ and the maximum potential value, a proxy for the size of the overdensity in figure \ref{fig:Qgrowth}.

We can see by looking at equations \eqref{FME_eqn}-\eqref{dba_dt} that at early times, $t \ll t_d$ and $\braket{\delta \hat a_i \delta \hat a_j}, \, \braket{\delta \hat a_i^\dagger \delta \hat a_j} \ll 1$, the evolution of the second order moments is governed by the $\sum_{kp} \Lambda^{ij}_{pl} \braket{\hat a_k} \braket{\hat a_p}$ term in equation \eqref{daa_dt}, i.e. at early times

\begin{align}
     \partial_t \braket{\delta \hat a_i \delta \hat a_j} \sim -i \sum_{kp} \Lambda^{ij}_{pl} \braket{\hat a_k} \braket{\hat a_p} \, .
\end{align}

This term is proportional to $[\hat a^\dagger, \hat a]$ and seeds the variance term, which is initially $0$ for a coherent state. We can take a second temporal derivative of equation \eqref{dba_dt} and plug in the above equation which should dominate at these early times, this implies that the initial evolution goes as 

\begin{align} \label{DefinitionKappa}
     \partial_{tt} \braket{\delta \hat a_i^\dagger \delta \hat a_j} &\sim 2 \mathbb{R}\left[ \sum_{kplbc} \Lambda^{ij}_{pl}  \Lambda^{kj}_{bc} \braket{\hat a_b} \braket{\hat a_c} \braket{\hat a_p^\dagger} \braket{\hat a_l^\dagger} \right]  \nonumber \\
     &\equiv \kappa_{ij} \, .
\end{align}

We now can define the quantity $\kappa$ which describes the rate of the growth of quantum corrections at early times. We can see in figure \ref{fig:Qgrowth} that initially $Q(t) = \mathrm{Tr}[\kappa_{ij}] \, t^2 / 2n_{tot}$. The approximation fits quite well at early times when the overdensity is growing slowly.  

Following this initial quadratic growth phase, the overdensity begins to grow quickly as the system collapses. During this collapse the wavefunction spreads exponentially around its mean field value. This continues until shell crossing, here defined to occur at $t = t_d$. The exponential spread of the wavefunction is behavior typical of quantum systems which exhibit classical chaos and corroborates previous studies of nonlinear systems, see for example \cite{eberhardt2021Q}. It is interesting to note that during this and the quadratic growth phase of $Q(t)$ the classical field behavior is still well approximated by classical particles, see for review \cite{Eberhardt2020}. We can see in figure \ref{fig:Qcompare} that the evolution of $Q(t)$ during the quadratic and exponential growth phases depend only weakly on the value of $\tilde \hbar$ and $n_{tot}$, assuming that both are chosen such that $Q(t) \ll 1$ during these phases. After shell crossing, the growth again becomes a power law for a time before experiencing runaway growth. 

\subsection{Effect of corrections on densities}

We plot the second order central moments as well as the difference in the spatial over-density given by the number operator, which is a second order moment, and the amplitude of the classical field in figure \ref{fig:varCompare}. We see that in the spatial case the covariance is predominantly off diagonal and therefore, there is no difference between the square amplitude of the spatial classical field and the spatial number operator. On the other hand the momentum density admits differences between $| a^{cl}(k_i)|^2$ and $\braket{\hat a^\dagger(k_i) \hat a(k_i)}$. We can see that a leading order effect of the corrections is to smooth interference patterns in the momentum density. This corroborates our intuition as it has been shown in previous work that phase diffusion is a leading order effect of quantum corrections, see for example figure 1 in \cite{eberhardt2021}.

\subsection{Behavior of the quantum breaktime}

We simulate a number of systems with different occupations, $n_{tot}$, and values of $\tilde \hbar$, and then plot the breaktime as defined in equation \eqref{breakTime} in figure \ref{fig:sinCollapse_tbr}. We see a logarithmic scaling of the breaktime during the collapses phase and then a power law scaling following shell crossing, a logarithmic enhancement in the breaktime is consistent with the expected behavior of quantum systems which exhibit classical chaos \cite{Hertberg2016, eberhardt2021Q}. This behavior is consistent across all the systems we tested. We can see that systems with large occupations at fixed $\tilde \hbar$, as well as larger $\tilde \hbar$ at fixed $n_{tot}$, have longer breaktimes. These results align with our expectation as it is generally argued that large occupation number per deBroglie wavelength imply long quantum breaktimes \cite{Hertberg2016,Guth2015}. 

The system with the largest value of $\tilde \hbar$  we tested, the blue line in figure \ref{fig:sinCollapse_tbr}, had all the dynamical momentum modes highly occupied even at the breaktime. This is the system that behaved most similar to the classical field theory. We can see that the slope of the logarithmic scaling of the breaktime in figure \ref{fig:sinCollapse_tbr} is approximately $1/7$. This means that even as $Q(t)$ was no longer $\ll 1$ its growth was still well described by the exponential growth shown in figure \ref{fig:Qcompare}. 

A typical realistic system would be continuously undergoing collapsing and merging. From our results studying the spread of the wavefunction we can then expect that during these processes the wavefunction spreads exponentially around its mean field value but spread slowly during periods of linear growth or once the system is already virialized. We expect quantum effects to start becoming very large when $Q(t) \sim 1$. Given the growth rates we found in figure \ref{fig:Qcompare} and assuming a system that is constantly undergoing collapse and merging events, this would imply that this happens when $e^{7t} \sim n_{tot}$, at such high occupations this time depends only weakly on the initial quadratic growth of $Q(t)$. This would imply a breaktime 

\begin{align} \label{TbreakEstimate}
    t_{br} \sim \frac{\ln(n_{tot})}{7} \, t_d \, ,
\end{align}

in the large $n_{tot}$ limit. This is about $\sim 30$ dynamical times at $n_{tot} \sim 10^{100}$. However, it is important to note that the wavefunction is only exponentially growing during nonlinear growth. We found that when the potential is slowly changing or the system has already collapsed the wavefunction only spreads slowly around its mean field value. 

Squeezing, the shrinking of the spread of the wave function in one direction in phase space, is accompanied by the initial quadratic spreading of the wave function $Q(t) = \mathrm{Tr}[\kappa_{ij}] \, t^2 / 2n_{tot}$, see equation \eqref{DefinitionKappa}. 
The time scale at which this genuine quantum effect because large, the so-called squeezing time scale, was discussed in previous work \cite{kopp2021nonclassicality,eberhardt2021Q}.  However, as we already noted, $t_{br}$ is dominated by the exponential growth which is due to the chaotic behavior. 
It thus not expected that the squeezing time scale is related to the quantum break time \eqref{TbreakEstimate}. 
The squeezing time scale was derived using the single mode (``Hartree'') ansatz in \cite{kopp2021nonclassicality}. However it was shown in \cite{eberhardt2021Q} that in multimode system deviations occur already before the single-mode squeezing time scale. Here we shown the existence a physically distinct time scale at which quantum effects become strong, $t_{br}$ in equation \eqref{TbreakEstimate}, which is genuine multi-mode phenomenon.

\section{Conclusions} \label{sec:conclusions}

We have simulated the gravitational collapse of an initial overdensity with long range attractive interactions tracking the mean and second order field operators. The second order operators tell us how the quantum field is spreading around its mean field value. We find that this spread is exponential during the nonlinear growth and collapse of the overdensity, see figure \ref{fig:Qgrowth}. However, at very early times or after the collapse the wavefunction spreads much slower, growing only as a power law. Interestingly we find that the rate of exponential growth depends only weakly on the mass of the constituent particles as long as we remain in the highly occupied regime, see figure \ref{fig:Qcompare}.

Using our definition of the quantum breaktime in equation \eqref{breakTime}, we can determine an approximate timescale \eqref{TbreakEstimate} when quantum corrections to the classical field theory evolution are no longer small. This is done by measuring how the uncertainty compares with the mean field value. The breaktime scales logarithmically with $n_{tot}$ for a time and then shifts to a power law scaling, see figure \ref{fig:sinCollapse_tbr}. A logarithmic enhancement in the quantum breaktime with increasing total particle number is typical of chaotic systems and corraborates earlier results studying similar problems \cite{eberhardt2021Q,Hertberg2016, Han2016,Albrecht2014}.

We have found that nonlinear growth drives the exponential spread of the wavefunction. This work then implies that quantum corrections are most significant for systems which have histories of continuous and violent nonlinear growth and mergers. For highly occupied systems the breaktime depends very weakly on the spread of the wavefunction at early times. We can therefore put a a rough order of magnitude estimate on the timescale when quantum corrections grow large as $t_{br} \sim \frac{\ln(n_{tot})}{7} \, t_d$. The tendency of the quantum corrections to decay field amplitudes and smooth oscillation peaks implies that they may be most important for phenomena that depend strongly on the interference properties of SFDM evolution such as \cite{Dalal2022}.

It is important to note the following limitations of this analysis and the need for future work addressing them. First, the test problems we simulate here are very simple and include only a single spatial dimension. More realistic initial conditions in three spatial dimensions with an expanding background would produce a more reliable estimation of the breaktime. Second, our approximation scheme tracks only the first two moments of the field, effectively a Gaussian approximation. This is the lower order correction to the mean field theory and including the contribution of higher order moments may alter the prediction of the quantum breaktime. 3D simulations which include higher order moment contributions will be included in a subsequent work. Finally, we have neglected to include the effect of decoherence by the environment which has the tendency to prevent the formation of observable macroscopic superpositions. Understanding this effect requires both an estimation of the decoherence timescale as a function of mass, as is done in \cite{Allali_2020, Allali_2021}, and the pointer states into which the quantum state is effectively projected. 

\begin{figure}[!ht]
	\includegraphics[width = .45\textwidth]{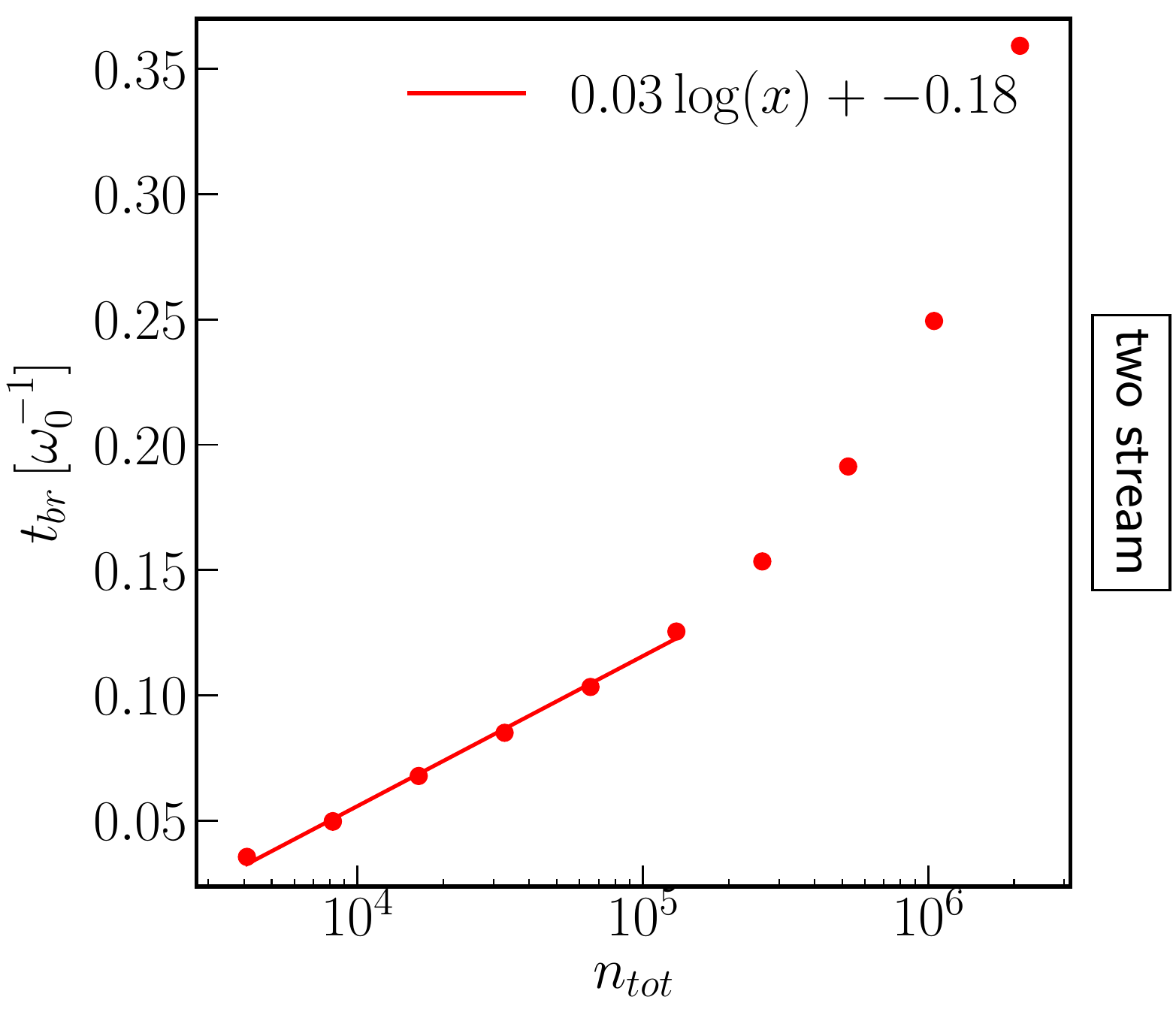}
	\caption{ Here we show the scaling of the breaktime with $n_{tot}$ for two stream instability. Like in case of the gravitational overdensity we see an initial logarithmic scaling followed by a power law scaling. The logarithmic scaling occurs during the time when the potential is growing.  }
	\label{fig:sinCollapse_tbr_2s}
\end{figure}

\begin{figure}[!ht]
	\includegraphics[width = .45\textwidth]{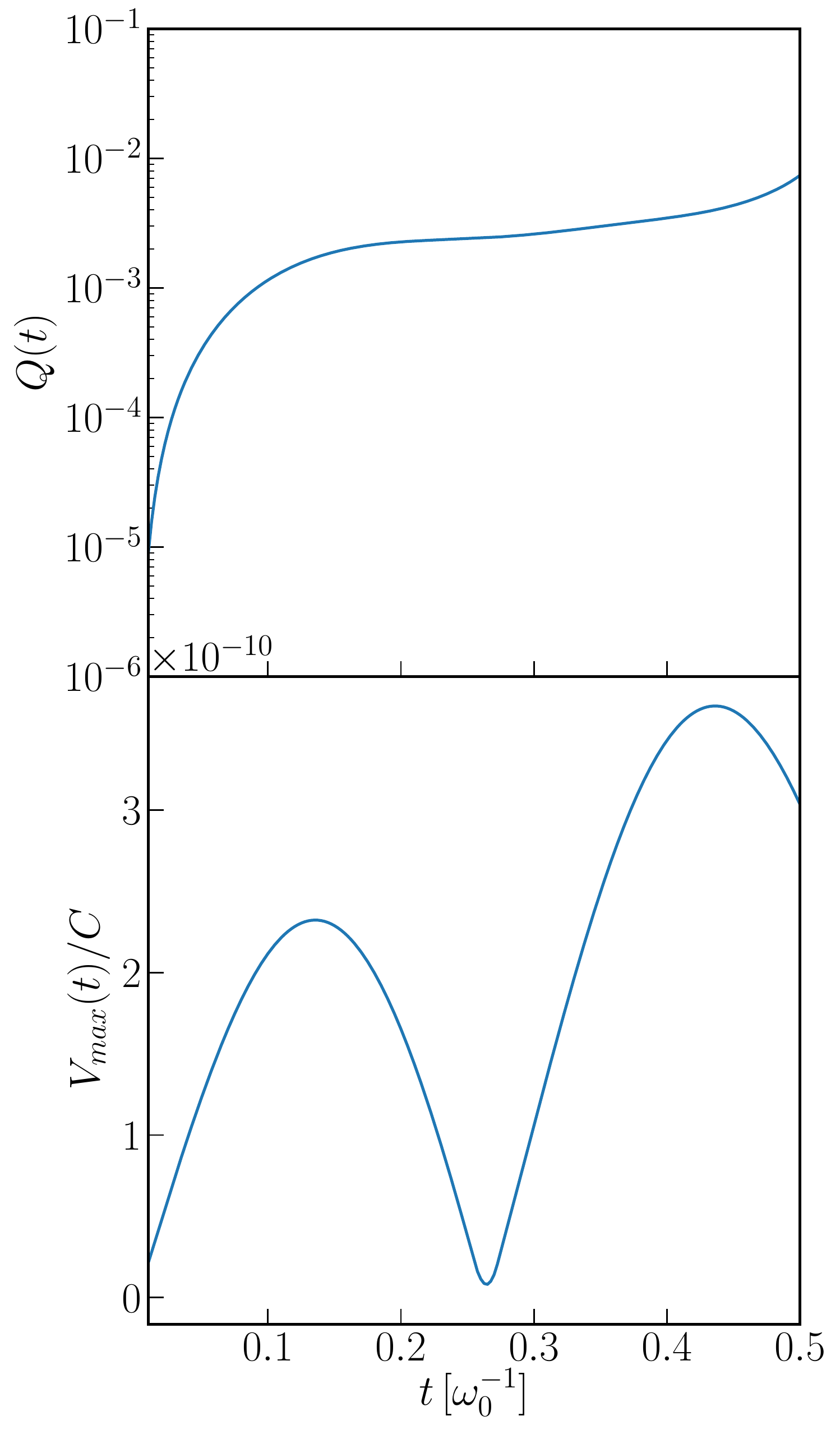}
	\caption{ Here the top plot shows the evolution of $Q$ and the bottom plot the maximum value of the potential, $V$, for the gravitational collapse of a sinewave overdensity. Here we set $\tilde \hbar = 2.5 \times 10^{-4}$, $n_{tot} \approx 10^{6}$. }
	\label{fig:twoStreamoverdense}
\end{figure}

\begin{figure}
	\includegraphics[width = .45\textwidth]{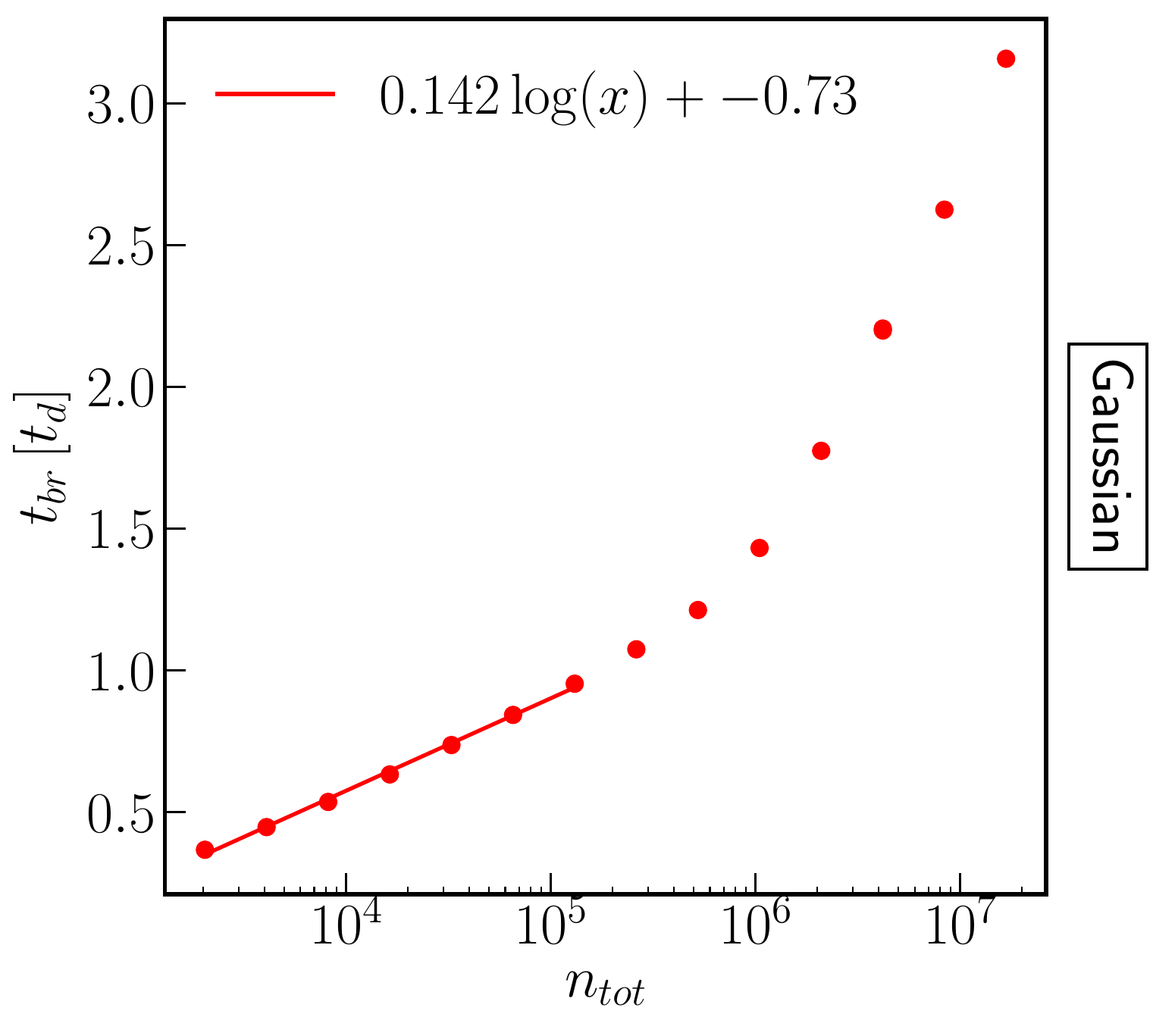}
	\caption{ Here we show the scaling of the breaktime with $n_{tot}$ for the gravitational collapse of a Gaussian. Like in case of the gravitational overdensity we see an initial logarithmic scaling followed by a power law scaling. The logarithmic scaling occurs prior to shell crossing.  }
	\label{fig:gaussCollapse_tbr}
\end{figure}

\appendix*
\appendix*
\renewcommand{\thesubsection}{\Alph{subsection}}
\section{Appendices}
\subsection{Other test systems} \label{appendix}

We also test other initial conditions and interactions which largely corroborate the results we have already demonstrated.

We simulate two counter propagating uniform density streams with an initial velocity perturbation. This system is relevant to the evolution of cold plasmas and is a common test problem \cite{Infeld_1969, thorne, Eberhardt2020}. The initial field is given by

\begin{align}
    \psi(x) = \sqrt{\frac{n_{tot}}{2}} {e^{-i \delta v L \cos(2 \pi x / L) / 2 \pi \tilde \hbar}}\left( e^{i x v_0 / \tilde \hbar} + e^{-i x v_0 / \tilde \hbar} \right) \, .
\end{align}

$\delta v$ is the initial velocity perturbation and $v_0$ the initial velocity of the streams. If the velocity separation, $2 v_0$, is less than the critical velocity associated with the perturbation, $v_c = \sqrt{C/2 \pi}$, then the perturbation will be unstable and grow exponentially, \cite{thorne}. Unlike the gravitational case, a single-stream initial condition would be stable for the electric case. Therefore we consider the two-stream perturbation which is unstable when $C > 0$, representing a repulsive long range interaction.

As for the previous test problem we evolve the system and track the spread of the wavefunction until the breaktime, see figure \ref{fig:sinCollapse_tbr_2s}. We see that the uncertainty grows in a similar way. There is an initial phase of power law growth followed by a phase of exponential growth and then a final phase of power law growth. We can see that this transition occurs when the potential stops growing, see figure \ref{fig:twoStreamoverdense}. This again corroborates the idea that it is the nonlinear growth of the overdensity that is cause the wavefunction to spread exponentially around its mean field value.

We also tested the gravitational collapse of a Gaussian with initial field given by 

\begin{align}
    \psi(x) = e^{-x^2/4 \sigma^2} / \,  \mathrm{Norm} \, .
\end{align}

The field is normalized such that $\sum_x |\psi(x)|^2 = n_{tot}$. Like the sinewave collapse, we find that the breaktime scales logarithmically prior to shell crossing and then as a power law, see figure \ref{fig:gaussCollapse_tbr}. 

We can see in these test problems the same qualatative behavior as in the sine wave collapse case. Specifically, logarithmic scaling of the breaktime when the potential is quickly growing and power law growth at other times. This general behavior corraborates the idea that it is the nonlinear growth that drives the exponential spread of the wavefunction.

\begin{figure}
	\includegraphics[width = .45\textwidth]{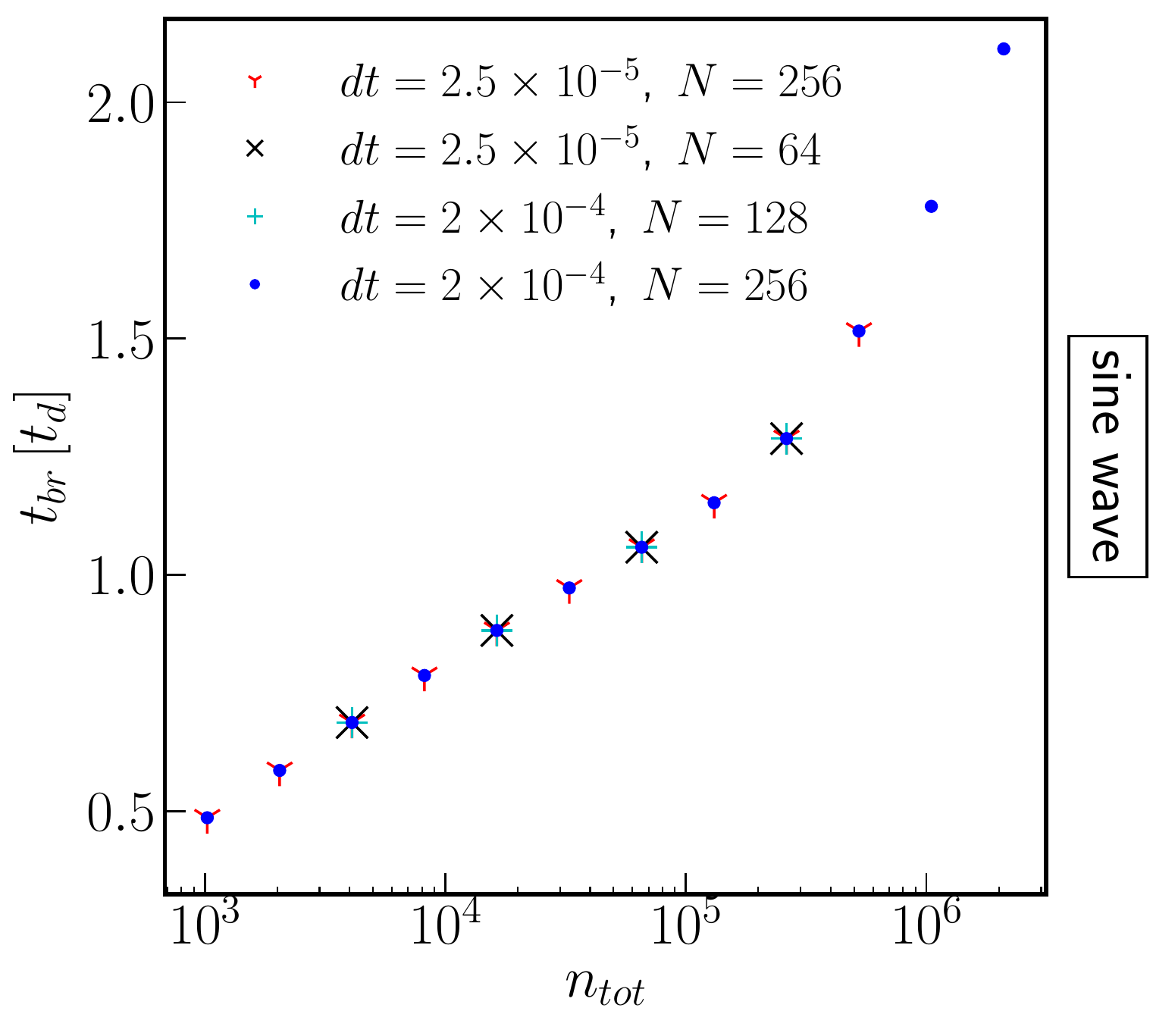}
	\caption{ Here we plot the breaktime as a function of $n_{tot}$ for an initial sine wave overdensity with $\tilde \hbar = 2 \times 10^{-3}$. We include points with a variety of $dt$ and $N$, we can see that the fractional difference between points at different temporal or spatial resolutions is very small with all point depending only on the total particle number $n_{tot}$. This demonstrates that our solutions are converged at the resolution used.}
	\label{fig:numConverge}
\end{figure}

\subsection{Numerical convergence tests} \label{appendixB}

Some simulation parameters are chosen which do not explicitly depend on the physics of the problem but instead reflect limitations in numerical resources. It is important to demonstrate that the results of our simulation do not depend on the choice of these parameters. In this appendix we will look at the timestep $dt$ and the mode number $N$. In both instances we will show that varying the value of these parameters does not change our result and we can therefore consider our results to be converged.

We run simulations with $\tilde \hbar = 2 \times 10^{-3}$ for $n_{tot} \in [10^3, 10^6]$ and a variety of $dt$ and $N$. We plot the results in figure \ref{fig:numConverge}. 
It is clear from the figure that our breaktime results are converged and do not depend on the temporal or spatial resolution used in the simulation. 

Because the field moment solver uses the same set of modes present in the classical field simulation it is subject to the same constraint on the choice of mode number, namely that the maximum acheived velocity is less than $\pi \tilde \hbar / dx$. The constraint on the choice of time step requires that the time derivative is small compared to any of the moments updated, namely that $dt \partial_t M / M \ll 1$. These conditions are discussed at length in the classical field case in \cite{Eberhardt2020}. In this appendix we show the results for a specific value of $\tilde \hbar$, but the temporal and spatial convergence was checked for all simulation results presented. 

\bibliography{BIB}

\end{document}